\documentclass{emulateapj}
\usepackage{epstopdf}

\newcommand{\kms}{km~s$^{-1}$}
\newcommand{\msun}{M$_{\sun}$}

\newcommand{\vsini}{$V_{eq}\sin{i}$}
\newcommand{\lbol}{$\log_{10}{L_{bol}/L_{\sun}}$}

\newcommand{\lii}{\ion{Li}{1}}

\newcommand{\name}{2MASS~J03202839$-$0446358}
\newcommand{\namesh}{2MASS~J0320$-$0446}

\slugcomment{Submitted to AJ 19 June 2008; accepted 19 March 2009}

\shorttitle{Age of 2MASS 0320-0446AB}
\shortauthors{Burgasser \& Blake}

\begin{document}

\title{An Age Constraint for the Very Low-Mass Stellar/Brown Dwarf Binary 2MASS~J03202839$-$0446358AB}

\author{Adam J.\ Burgasser}
\affil{Massachusetts Institute of Technology, Kavli Institute for Astrophysics and Space Research,
Building 37, Room 664B, 77 Massachusetts Avenue, Cambridge, MA 02139, USA; ajb@mit.edu}
\and
\author{Cullen H.\ Blake\altaffilmark{1} }
\affil{Harvard-Smithsonian Center for Astrophysics, 60 Garden Street, Cambridge, MA 02138}

\altaffiltext{1}{Harvard Origins of Life Initiative Fellow}

\begin{abstract}
2MASS~J03202839$-$0446358AB is a recently identified, late-type M dwarf/T dwarf 
spectroscopic binary system for which both the radial velocity orbit for the primary
and spectral types for both components have been
determined.  By combining these measurements with predictions from 
four different sets of evolutionary models, we determine a 
minimum age of 2.0$\pm$0.3~Gyr for this system,  corresponding
to minimum primary and secondary masses of
0.080~{\msun} and 0.053~{\msun}, respectively.  We find broad agreement in 
the inferred age and mass constraints
between the evolutionary models, including those
that incorporate atmospheric condensate grain opacity; however, 
we are not able to independently assess their accuracy.
The inferred minimum age agrees with the kinematics
and absence of magnetic activity in this system, but not the rapid rotation of its
primary, further evidence of a breakdown in angular momentum evolution
trends amongst the lowest luminosity stars.
Assuming a maximum age of 10~Gyr, we
constrain the orbital inclination of this system to
$i \gtrsim 53\degr$. 
More precise constraints on the orbital inclination and/or component masses of 
2MASS~0320$-$0446AB, through either measurement of the secondary radial
velocity orbit (optimally in the 1.2--1.3~$\micron$ band) or detection of an eclipse (only 0.3\% probability based on geometric constraints),
would yield a bounded age estimate for this system, and the opportunity to use it as an empirical test for brown dwarf evolutionary models at late ages.
\end{abstract}


\section{Introduction}

Of the three most fundamental parameters of a star---mass, age and composition---age is arguably
the most difficult to obtain an accurate measure.  
Direct measurements of mass
(e.g., orbital motion, microlensing, asteroseismology) and atmospheric 
composition (e.g., spectral analysis) are possible for individual stars, but
age determinations are generally limited to the coeval stellar systems for which
stellar evolutionary effects can be exploited (e.g., pre-main sequence contraction, isochronal ages, post-main sequence turnoff).  
Individual stars can be approximately age-dated using
empirical trends in magnetic activity, element depletion, rotation or kinematics that are calibrated against cluster populations and/or numerical simulations (e.g., \citealt{wil70,sku72,wie77,bar07,mam08,wes08}).
However, such trends are fundamentally statistical in nature, and 
source-to-source scatter can be comparable in magnitude to
mean values.
Age uncertainties are even more problematic for the lowest-mass stars (M $\lesssim$ 0.5~{\msun}), as
post-main sequence evolution for these objects occurs at ages much greater
than a Hubble time, and activity and rotation trends present in solar-type stars
begin to break down (e.g., \citealt{moh03,rei08}). For the vast majority
of intermediate-aged (1--10~Gyr), very low-mass stars in the Galactic disk, 
barring a few special
cases (e.g., low-mass companions to cooling white dwarfs; \citealt{sil06})
age determinations are difficult to obtain and highly uncertain.

Ages are of particular importance for even lower-mass brown dwarfs (M $\lesssim$ 0.075~{\msun}), objects which fail to sustain core hydrogen fusion and therefore
cool and dim over time
\citep{kum62,hay63}.  The cooling rate of a brown dwarf is set by its age-dependent luminosity, while its initial reservoir of thermal energy is set by gravitational contraction and hence total mass.  As such, there is an inherent degeneracy between the mass, age and observable properties of a given
brown dwarf in the Galactic field population; one cannot distinguish between a young, low-mass brown dwarf and an old, massive one from spectral type, luminosity or effective temperature alone.
This degeneracy can be resolved for individual sources 
through measurement of a secondary parameter such as surface gravity, 
which may then be compared to predictions from brown dwarf evolutionary models
(e.g., \citealt{moh04,metgrav,leg07}).
However, surface gravity determinations are highly dependent on the accuracy of
atmospheric models, which are known to have systematic problems
at low temperatures due to incompleteness in molecular opacities (e.g., \citealt{fre08})
and dynamic atmospheric processes (e.g., \citealt{cus08}). 
Discrete metrics such as the presence of absence of {\lii} absorption (depleted in brown dwarfs more
massive than 0.065~{\msun} at ages $\gtrsim$200~Myr \citealt{reb92,bil97}), are generally
more robust but do not provide a continuous measure of age for brown dwarfs in the Galactic field population.

Binary systems containing brown dwarf components can be used to break this mass/age 
degeneracy without resorting to atmospheric models.  Specifically, systems
for which masses can be determined via astrometric and/or spectroscopic orbit measurements,
and component spectral types, effective temperatures and/or luminosities assessed,
can be compared directly with evolutionary models to uniquely constrain
the system age (e.g., \citealt{liu08}).  Furthermore, by comparing the inferred ages and masses for each presumably coeval component, such systems can provide empirical tests of the evolutionary models themselves.
A benchmark example
is the young ($\sim$300~Myr) binary---and perhaps triple---brown dwarf
system Gliese 569B
\citep{mar00,lan01,zap04,zap05,sim06}.  With both astrometric and
spectroscopic orbit determinations, and resolved component spectroscopy,
this system has been used to explicitly test evolutionary model tracks
and lithium burning timescales \citep{zap04,zap05} as well as derive component ages,
which are found to agree qualitatively with kinematic arguments (e.g., \citealt{ken01}).  Other close binaries
with astrometric or spectroscopic orbits have also been used for direct mass 
determinations (e.g., \citealt{bou04,sta06,joe07,ire08,liu08}), but these
systems generally lack resolved spectroscopy and therefore precise component characterization.  
They have also tended to be young, preventing stringent tests of the
long-term evolution of cooling brown dwarfs.
Older, nearby very low-mass binaries with resolved spectra (e.g., \citealt{me2200,mce1707,mar06,stu09}) generally have prohibitively long orbital periods
for mass determinations.

Recently, we identified a very low-mass binary system
for which a spectroscopic orbit and component spectral
types could be determined: the late-type source {\name} (hereafter {\namesh}; \citealt{cru03,wil03}).  Our independent discoveries of this system were made via two complementary techniques.  
\citet[hereafter Bl08]{bla08} identified this source as a single-lined
radial velocity variable, with a period of 0.67~yr and separation $\sim$0.4~AU,
following roughly 3 years of high-resolution, near-infrared 
spectroscopic monitoring
(see \citealt{bla07}).
\citet[hereafter Bu08]{me0320} demonstrated that the 
near-infrared spectrum of this source could be reproduced as
an M8.5 plus T5$\pm$1 unresolved pair, based on the spectral template matching technique
outlined in \citet{me0805}.  
The methods used by these studies have yielded both 
mass and spectral type constraints for the components
of {\namesh}, and thus a rare
opportunity to robustly constrain the age
of a relatively old low-mass star and brown dwarf system
in the Galactic disk. 

In this article, we determine a lower limit for the age of {\namesh} by combining the
radial velocity measurements of Bl08 and component spectral type determinations of Bu08 with
current evolutionary models.  Our method is described in $\S$~2, which includes discussion of sources of empirical uncertainty and systematic variations from four
sets of evolutionary models.  We obtain lower 
limits on the age, component masses, and orbital inclination of the system, and compare our age constraint to expectations based on kinematics, magnetic activity and rotation of the primary component.  In $\S$~3 we discuss our results, focusing in particular on 
how future observations could provide bounded limits on the age and component masses of this system, and thereby facilitate tests
of the evolutionary models themselves at late ages.

\section{The Age of {\namesh}}

\subsection{Component Luminosities}

Evolutionary models predict the luminosities and effective temperatures
of cooling brown dwarfs over time, parameters that have been shown to 
correlate well with spectral type (e.g., \citealt{kir00,gol04,nak04}). 
Luminosity is the more reliable parameter, being based on the measured
distance and broad-band spectral flux of a source, as opposed to model-dependent determinations of
photospheric gas temperature and/or radius.  However, in the case
of {\namesh}, neither distance nor component fluxes have been measured, the latter due to the fact that this system is as yet unresolved (and for the near future, unresolvable; see $\S$~3).  We therefore used the 
component spectral types of this system and luminosity measurements
for similarly-classified, single (unresolved) sources from \citet{dah02,gol04,vrb04} and \citet{cus05} to estimate the component luminosities.

For the M8.5 primary, there are 13 M8--M9  
field dwarfs with bolometric luminosities (parallax distance and broad-band spectral flux measurements) reported in the studies listed above.  
Two of the sources---the M8 LHS 2397a, a known binary \citep{fre03}; and the
M9 LP 944-20, believed to be a younger system ($\sim$500~Myr, \citealt{tin98})---are unusual sources
and therefore excluded from this analysis. The mean
bolometric magnitude of the remaining stars is 
${\langle}M_{bol}{\rangle}$ = 13.36$\pm$0.29 mag, corresponding to {\lbol} = -3.45$\pm$0.12.

For the T5$\pm$1 secondary, there are fewer field brown dwarfs 
with reliable luminosity measurements 
(1 T4.5 dwarf and 5 T6 dwarfs)
and these show considerably greater scatter in their bolometric
magnitudes: ${\langle}M_{bol}{\rangle}$ = 17.2$\pm$0.6.  This scatter may be due in part
to unresolved multiplicity, which appears to be enhanced amongst
the earliest-type T dwarfs \citep{mehst2,liu06,meltbinary}.  
Hence, we estimated the luminosity of {\namesh}B
using the $M_{bol}$/spectral type relation of 
\citet{meltbinary}\footnote{The coefficients of this polynomial relation reported
in \citet{meltbinary} did not list sufficient significant digits, resulting in slight
differences between the numerical relation and that shown in Figure~1 of the paper.  The coefficients
as defined should be \{$c_i$\} = [1.37376e1, 1.90250e-1, 1.73083e-2, 7.40013e-3, -1.75144e-3, 1.14234e-4, -2.32248e-06], where $M_{bol}$ = $\sum_{i=0}^{6}{c_i}$SpT$^i$ and SpT(T0) = 10, SpT(T5) = 15, etc.}.
A mean ${\langle}M_{bol}{\rangle}$ = 17.09$\pm$0.29 ({\lbol} = -4.94$\pm$0.17) was adopted,
where we have taken into account the uncertainty in the secondary spectral
type and the $M_{bol}$/spectral type relation (0.22~mag).  This value agrees well with estimates
from \citet[${\langle}M_{bol}{\rangle}$ = 16.9$\pm$0.4]{vrb04} and \citet[${\langle}M_{bol}{\rangle}$ = 17.3$\pm$0.6]{gol04}.

\subsection{Evolutionary Models}

In order to assess systematic uncertainties in the derived age and component properties, we considered four different sets of
evolutionary models in our analysis: the cloudless models of \citet[hereafter TUCSON models]{bur97,bur01}, the ``COND'' cloudless models of \citet[hereafter COND models]{bar03}, and the cloudless and cloudy models
from \citet[hereafter SM08 models]{sau08}.  All four sets of models assume solar metallicity, which is
appropriate given that composite red optical and near-infrared spectra of {\namesh} 
show no indications of subsolar metallicity \citep[Bu08]{cru03}.
The choice of ``cloudless'' evolutionary models (referring to the absence of condensate clouds in atmospheric opacities) is driven partly by their 
availability.  In addition, 
the spectral energy distributions of the M8.5 and T5 components of {\namesh}
are minimally affected by condensate cloud opacity (e.g., \citealt{all01}).  However, cloud opacity in the intermediate L dwarf stage may slow radiative cooling during this phase and bias the inferred age of the T-type
secondary (SM08), although \citet{cha00} have claimed that clouds have only a ``small effect'' on evolution.  To test this possibility, we chose to examine both the cloudless and cloudy SM08 models, the latter of which takes into account
photospheric cloud opacity in thermal evolution through the use of atmospheric models generated according to the prescriptions outlined in \citet{ack01} and SM08. 

\begin{figure}
\epsscale{1.0}
\plottwo{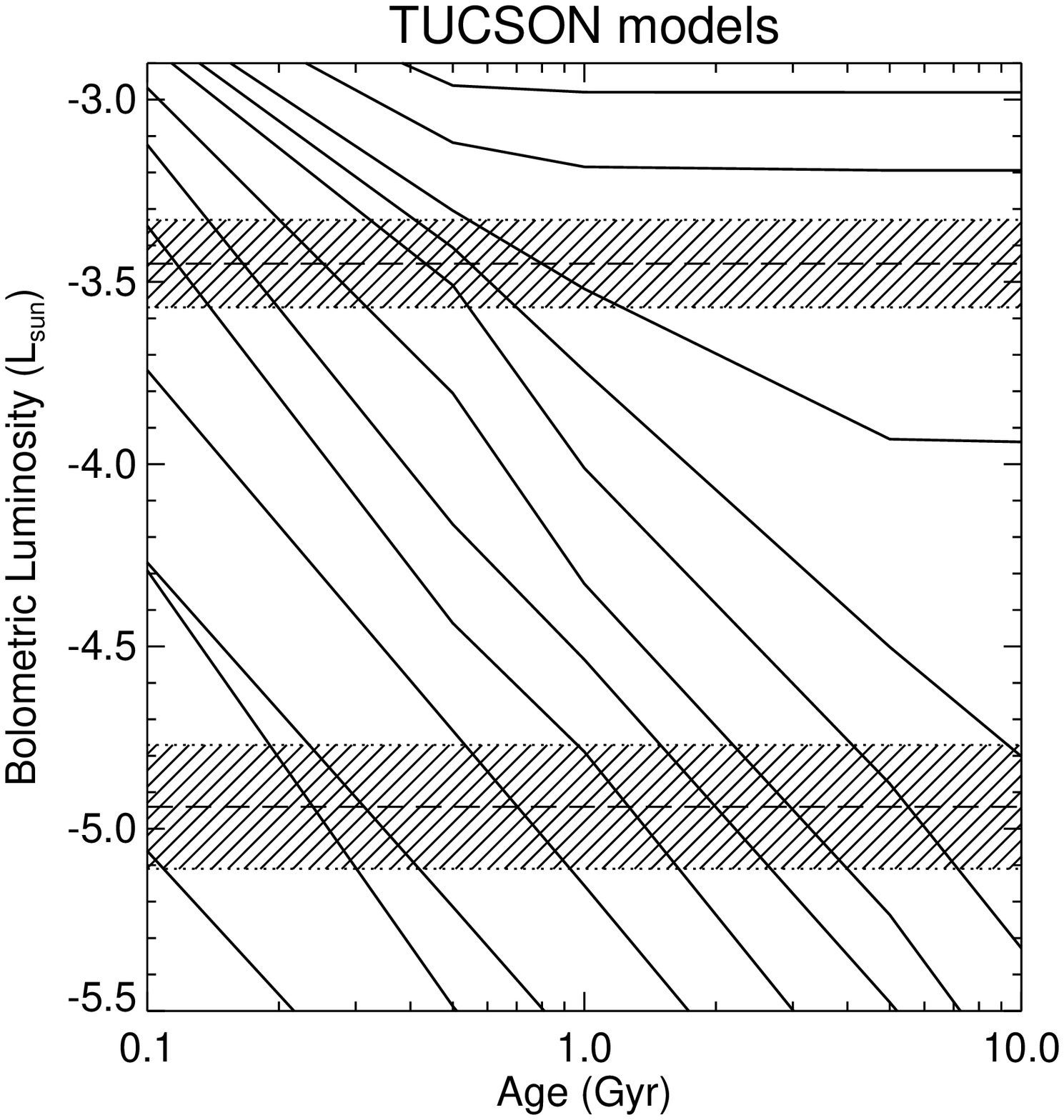}{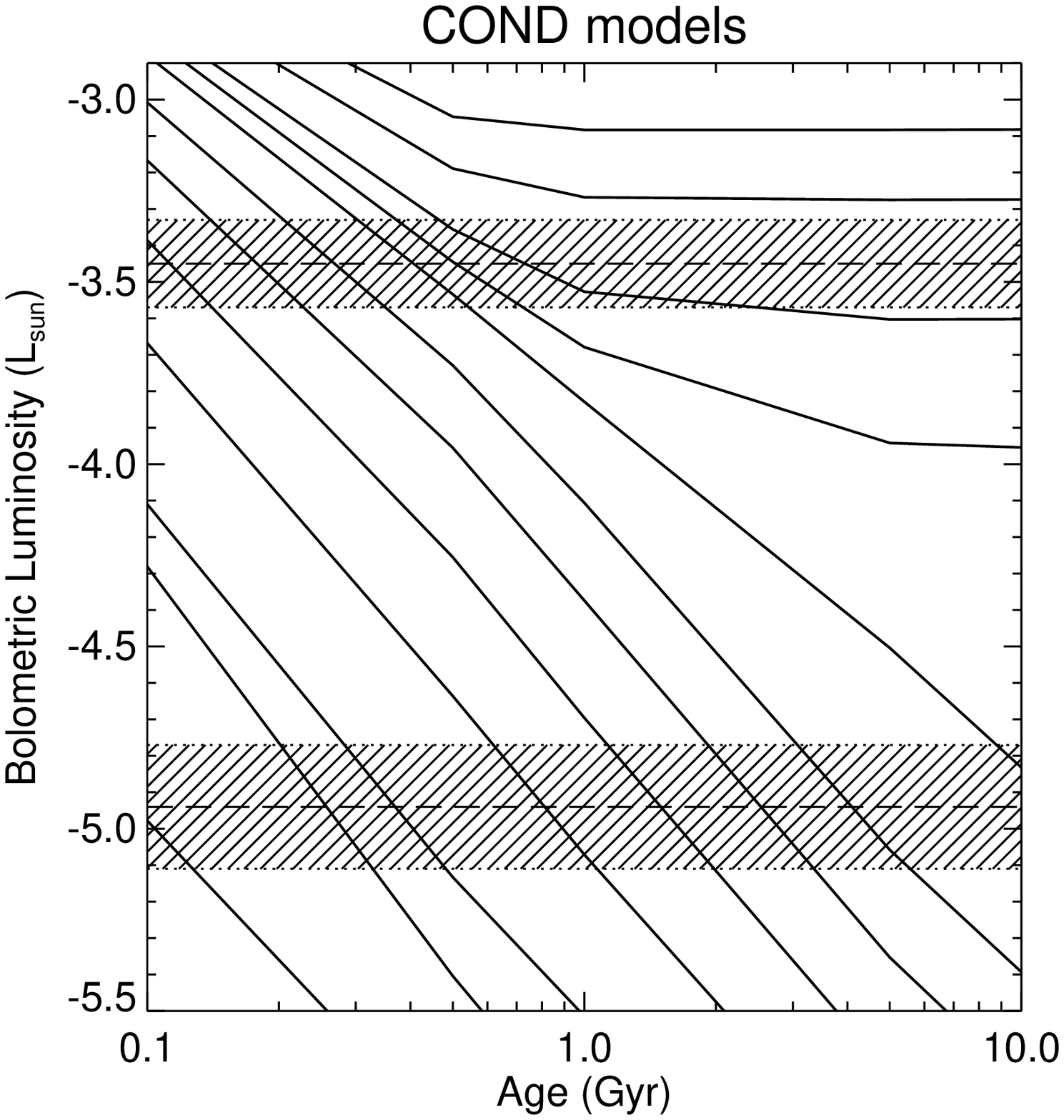} \\
\plottwo{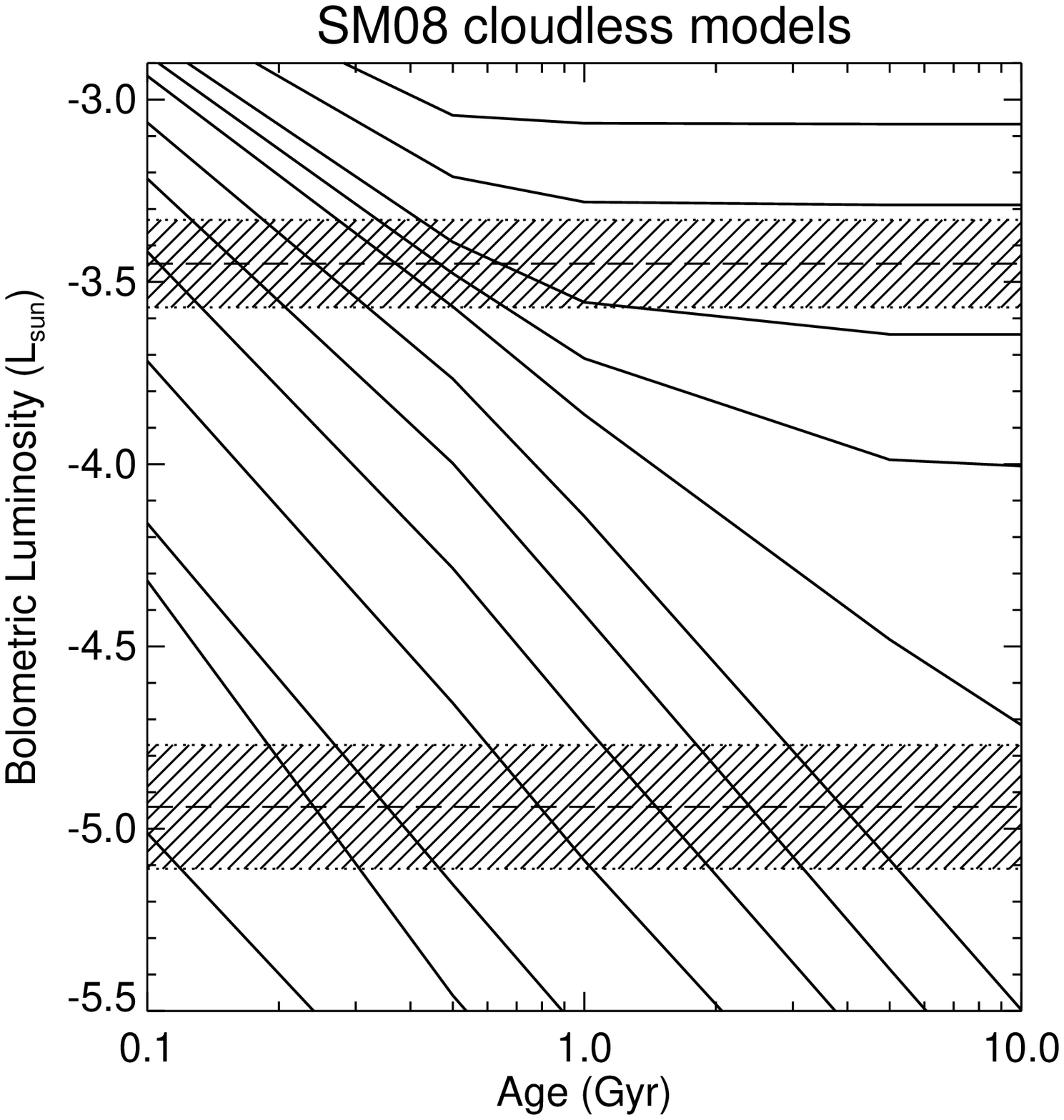}{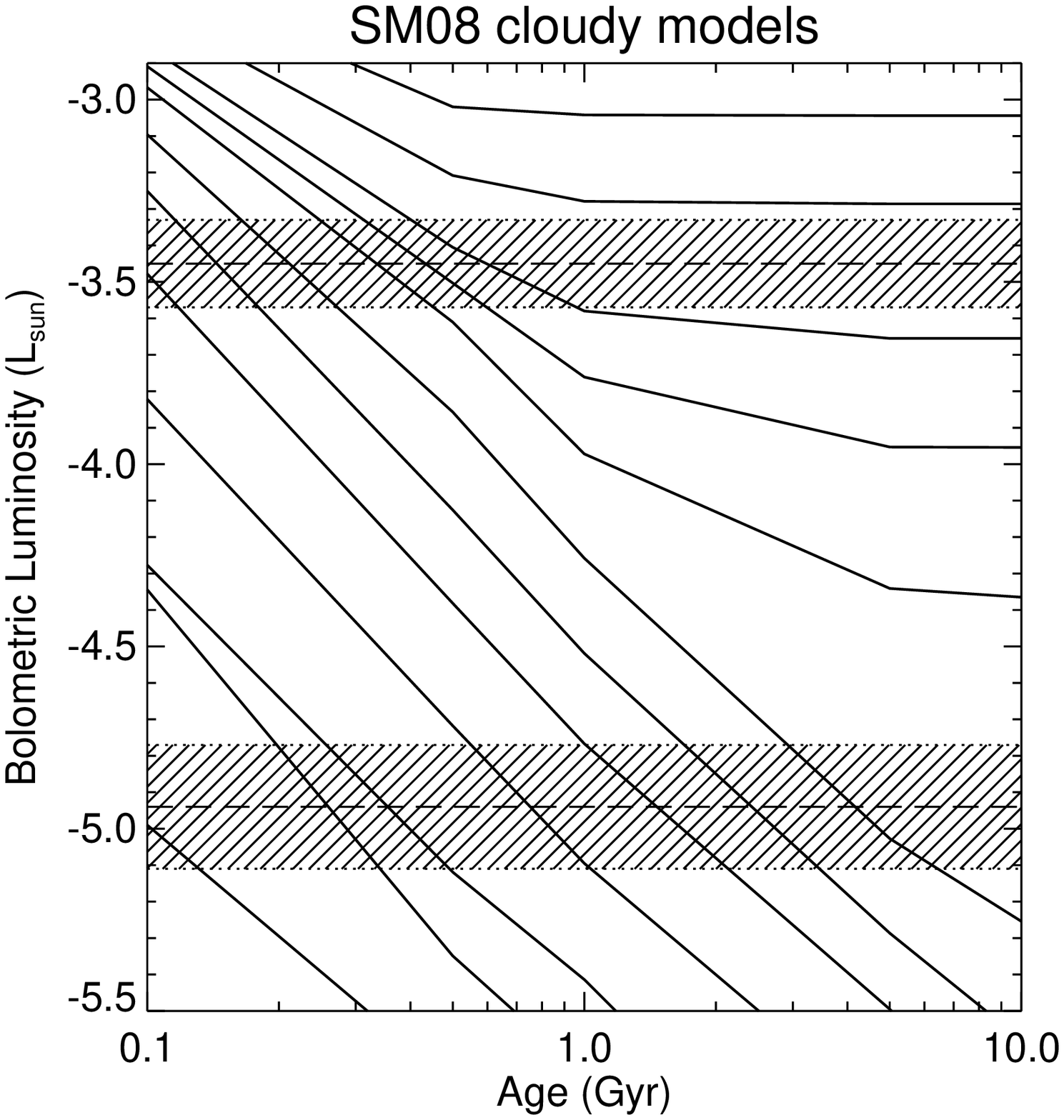}
\caption{Luminosity evolutionary tracks based on models from \citet[upper left]{bur97,bur01}; \citet[upper right]{bar03}; and \citet[bottom left: cloudless; bottom right: cloudy]{sau08}.  Tracks are shown for masses of 
0.01, 0.015, 0.02, 0.03, 0.04, 0.05, 0.06, 0.07, 0.075, 0.08, 0.09,
and 0.1~{\msun}, from lower left to upper right in each panel.  The estimated luminosities and uncertainties of the two components of {\namesh} are indicated by the hatched regions.
\label{fig_models}}
\end{figure}

\begin{figure}
\epsscale{1.0}
\plottwo{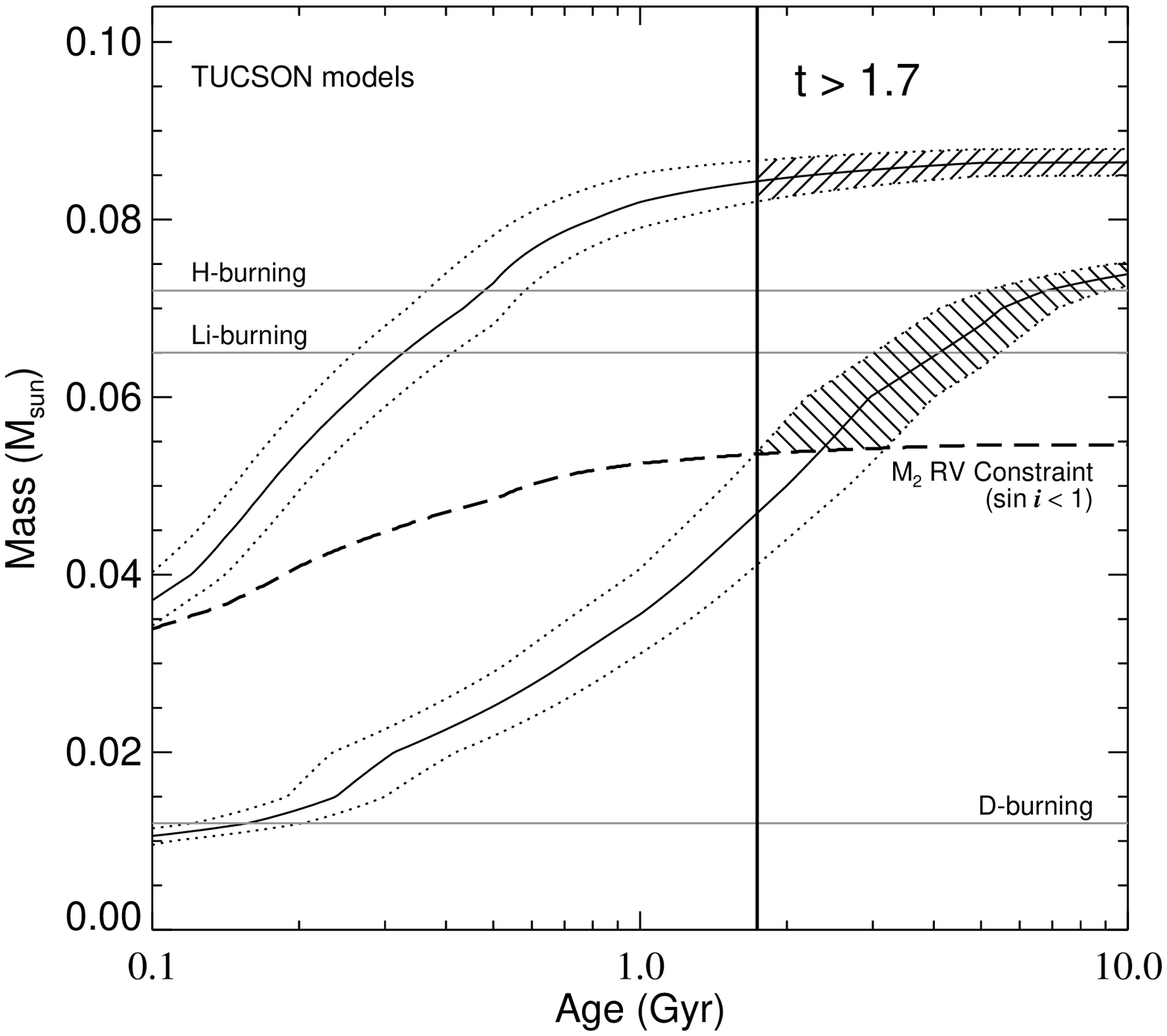}{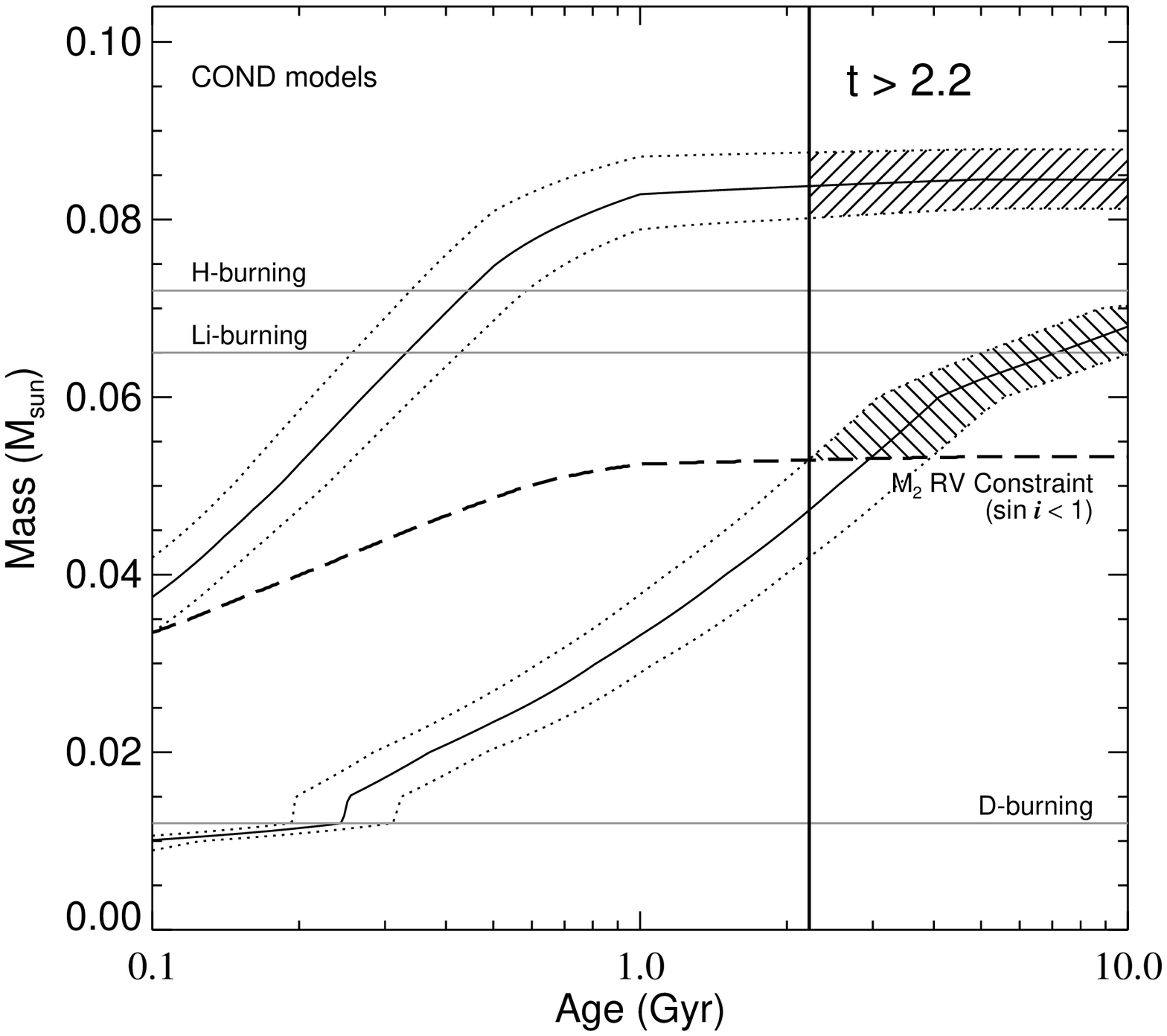} \\
\plottwo{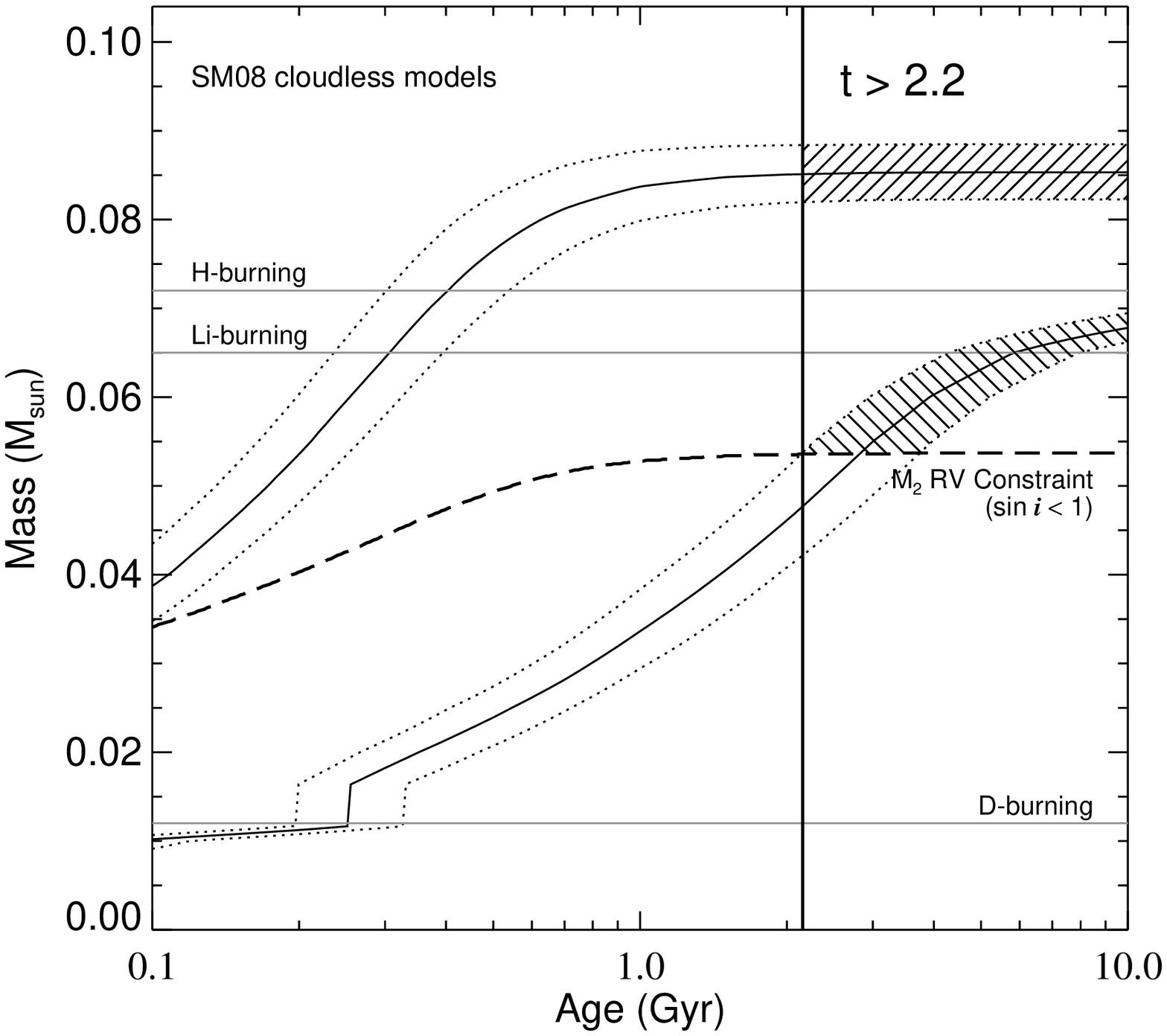}{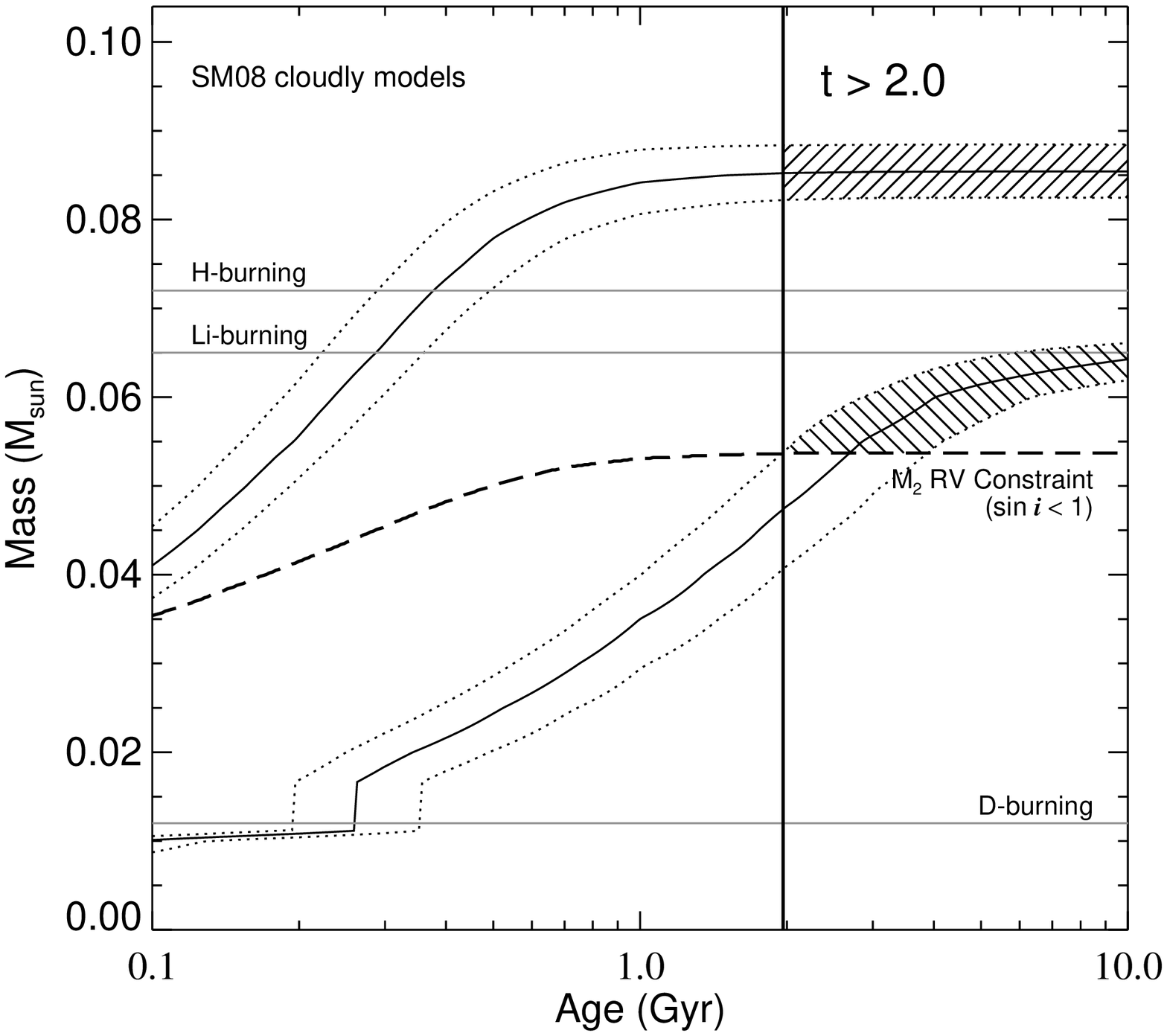}
\caption{{\namesh} component masses (solid lines; dotted lines indicate uncertainty range) as a function of age based on the component luminosity constraints shown in Figure~\ref{fig_models}.  
Age-dependent constraints on the minimum secondary mass based on the primary radial velocity orbit and lower bound of the primary mass, and assuming $\sin{i} \leq 1$, 
are indicated by dashed lines.  The intersection
of this line with secondary component mass constraints (based
on luminosity) sets the minimum age of the {\namesh} system,
1.7--2.2~Gyr.  The hatched regions indicate parameter spaces
allowed by the luminosity and radial velocity measurements and evolutionary
models.
Horizontal grey lines indicate threshold masses for core hydrogen (0.075~{\msun}), lithium (0.065~{\msun})
and deuterium fusion (0.013~{\msun}), assuming solar metallicity.
\label{fig_mvst}}
\end{figure}

Figure~\ref{fig_models} compares the luminosity estimates for the two components of {\namesh} to the evolutionary tracks of each model set.
The luminosities (and their uncertainties) constrain the mass/age parameter space of each component, as illustrated in Figure~\ref{fig_mvst}.
Component masses generally increase with system age, as more massive low mass stars and brown dwarfs take longer to radiate their greater reservoir of heat energy
from initial contraction.  The mass of the primary of this system reaches an asymptotic value of $\sim$0.08--0.09~{\msun} for ages $\gtrsim$1~Gyr, consistent with
a hydrogen-fusing very low-mass star.  If the system is younger than $\sim$400~Myr, the primary could be substellar.  Note that ages $\lesssim$300~Myr (primary masses $<$0.065~{\msun}) can be ruled out based on the absence of {\lii} absorption at 6708~{\AA} in the unresolved red optical spectrum 
of this source \citep{cru03}.
The mass of the secondary increases across the full age range shown in Figure~\ref{fig_mvst} as this component is substellar up to 10~Gyr.  There is some divergence in the evolutionary tracks at late ages for this component, however; the TUCSON models predict a mass near the hydrogen-burning limit, while the SM08 cloudless and cloudy models predict masses above and below the Li-burning minimum mass, respectively.
The kink in the mass/age relation of {\namesh}B at ages of 200-300~Myr, particularly in the COND and SM08 models, reflects the prolonged burning of deuterium in brown dwarfs with masses just above 0.013~{\msun}, producing higher luminosities at this temporary stage of evolution. The
mass ratio of the system, $q \equiv$ M$_2$/M$_1$, also increases
as a function of age,
ranging from $\sim$0.2 at 100~Myr to a maximum of $\sim$0.8 at 10~Gyr.

\subsection{Constraints from the Primary Radial Velocity Orbit}

With constraints in the mass/age phase space provided by the component luminosities and evolutionary
models, we can now use the radial velocity orbit to break the mass/age 
degeneracy. The radial velocity variations measured by Bl08 only probe the 
recoil velocity of the primary of the {\namesh} system.  These observations
provide a coupled constraint between the masses and inclination of the system:
\begin{equation}
{\rm M}_2\sin{i} = (0.2062{\pm}0.0034)({\rm M}_1+{\rm M}_2)^{2/3}~~{\rm M}_{\sun},
\end{equation}
(Bl08) where $i$ is the inclination angle of the orbit, and M$_1$ and M$_2$ are the masses of the primary and secondary components in solar mass units, respectively.
We can make a geometric constraint that $\sin{i} \leq 1$,
which yields a transcendental equation for the 
lower limit of the secondary component mass of the system
as a function of the primary component mass.  Using our age-dependent 
lower bound for the latter based on the evolutionary
models (including luminosity uncertainties) 
the constraint on ${\rm M}_2\sin{i}$ from Eqn.~1 translates into a  
minimum secondary mass as a function of age, as shown
in Figure~\ref{fig_mvst}. 
Finally, the age at which the upper bound of the secondary component
mass (based on the evolutionary models) crosses the radial velocity 
minimum mass line corresponds to the minimum age of the system.

All four models predict a minimum age for {\namesh} in the range
1.7--2.2~Gyr (Table~\ref{tab_modelfit}).
This age is in qualitative 
and quantitative agreement with those inferred by Bl08 from the kinematics of the
{\namesh} system\footnote{Combining the systemic radial velocity from Bl08 with the revised proper motion and distance estimates from Bu08, the $UVW$ space motions
of this system are $U$ = -38$\pm$5~{\kms}, $V$ = -20$\pm$3~{\kms} and $W$ = -32$\pm$4~{\kms}, where we assume an LSR solar motion of $U_{\sun}$ = 10~{\kms}, $V_{\sun}$ = 5.25~{\kms} and $W_{\sun}$ = 7.17~{\kms} \citep{deh98}. \citet{wie77}, equation 8, predicts an age $>$1.6~Gyr at the 95\% confidence level
for these kinematics.} and stellar age/activity trends.  In the case of the latter,
the optical spectrum of {\namesh}
shows no detectable H$\alpha$ emission
\citep{cru03}, even though $>$90\% of nearby 
M8--M9 dwarfs exhibit such emission \citep{giz00,sch07,wes08}. 
For comparison, \citet{wes08} 
estimate an ``activity lifetime'' (i.e., timescale for H$\alpha$ emission
to drop below detectable levels) of 8$^{+0.5}_{-1.0}$~Gyr
for M7 dwarfs.  This age may be too high of an estimate for {\namesh}, as the increase in activity lifetimes for spectral types M2--M7
observed by \citet{wes08} may not continue for later spectral types. 
Magnetic field lines are increasingly decoupled from lower-temperature photospheres, and the frequency and strength of H$\alpha$ emission decrease rapidly beyond type M7/M8
(e.g., \citealt{giz00,gel02,moh02,wes04}).  Hence, the absence of magnetic emission 
from {\namesh}A is merely indicative of an older age, as is its kinematics.  

Rotation
is a third commonly-used empirical age diagnostic for stars, based on the secular angular momentum loss observed amongst solar-type stars through the emission magnetized stellar winds (e.g., \citealt{sku72,sod91}).
However, so-called gyrochronological relations calibrated for these stars are not necessarily
applicable to lower-mass
objects, due both to the fully convective interiors of the latter and the decoupling of field lines from low-temperature atmospheres (e.g., \citealt{rei08}).   Indeed, {\namesh}A proves to be a relatively rapid
rotator, with an equatorial spin velocity of {\vsini} = 16.5$\pm$0.5~{\kms} and a rotation period $<$ 7.4~hr (Bl08).  For solar-type stars, this rapid rotation generally indicates a young age; an
extrapolation (in color) of gyrochronology relations by
\citet[equation 3, assuming $B-V$ $\approx$ 2.1 for {\namesh}A; e.g., \citealt{leg92}]{bar07} 
yields an age of only $\sim$0.1~Myr.  This is inconsistent with the absence of {\lii}
absorption, space kinematics and lack of magnetic emission from this source.  Clearly, rotation
does not provide a useful age metric for the {\namesh} system, further emphasizing the
breakdown of age/angular momentum trends in the lowest-mass stars.

The derived minimum age and radial velocity constraints
of the {\namesh} system allow us to
constrain model-dependent minimum masses for its components as well; M$_1$ $\geq$ 0.080--0.082~{\msun} and M$_2$ $\geq$ 0.053--0.054~{\msun}. The ranges in these values reflect variations between the four 
evolutionary models (Table~\ref{tab_modelfit}). Note that there is essentially no difference in the minimum masses inferred
from the cloudy and cloudless SM08 models.
The minimum mass ratio of the system, $q >$ 0.60--0.62, is also consistent across the evolutionary models, and in accord with the
general preference of large mass ratio systems observed amongst very low-mass 
binaries ($>$90\% of known binaries with $M_1 < 0.1$~{\msun}
have $q > 0.6$; see \citealt{me06ppv}).  

Finally, while the minimum ages and masses
of {\namesh} are inferred assuming $\sin{i} \leq 1$,
it is possible to constrain the maximum masses and minimum orbital inclination of this system assuming an upper age limit.  Adopting $\tau <$ 10~Gyr, we infer $\sin{i} >$ 0.80--0.86, corresponding to $i >$ 53$\degr$--59$\degr$.  This constraint is only slightly more restrictive than 
the $i > 44\degr$ lower limit determined by Bl08 assuming that the fainter
secondary must have a lower mass than the primary.  It does not 
significantly improve the chances that this system
is an eclipsing edge-on system ($\sim$0.3\% by geometry). Maximum primary (0.088~{\msun}) and secondary masses (0.066--0.075~{\msun}) are effectively set by the luminosity constraints and evolutionary models.  Here we see the
most significant difference between the models, a 13\% discrepancy 
in the maximum mass of the secondary component, likely due to different
treatments of light element fusion near the Li- and H-burning minimum
masses.  This variation confirms the importance of older binary systems as tests of long-term
brown dwarf evolution, particularly near fusion mass limits. We also note that there is
little difference ($\sim$4\%) in the maximum masses inferred from the SM08 cloudy and cloudless models, illustrating again the negligible role of cloud opacity in the long-term evolution of brown dwarfs like {\namesh}B.

\section{Discussion}

The combination of component luminosities, radial velocity
orbit of the primary and evolutionary models have allowed us to estimate the minimum
age of the {\namesh} system and its component masses.  
The ages are consistent between four evolutionary models of brown dwarfs, 
and are more precise (although not necessarily more accurate) than estimates based on
as-yet poorly-constrained statistical trends in kinematics, magnetic activity and angular momentum evolution.
However, a bounded age estimate remains elusive due to the unknown inclination of the system and determination of component masses.
As discussed in Bl08, the inclination of the {\namesh} orbit is irrelevant if the radial velocity orbit of the secondary
can be determined.  In that case, one need only compare the derived system mass ratio and component luminosities to evolutionary models to obtain a bounded age estimate.
Measurement of the secondary motion in the $K$-band data of Bl08 was not possible due to the very large flux contrast between the components; Bl08 rule out a contrast ratio of $\lesssim$10:1 at these wavelengths; spectral template fits from Bu08 predict a contrast ratio of $\gtrsim$350:1.  
A more effective approach would be
the acquisition of radial velocity measurements in the 1.2-1.3~$\micron$ band where the T dwarf secondary is considerably brighter and the contrast
ratio is closer to 20:1 (depending on the absolute magnitude scale; see discussion in Bu08).  At these contrasts, the radial velocity of the secondary can be measured
using existing techniques for high-contrast spectroscopic binary systems (e.g., \citealt{zuc94}).

Alternately, if this system is observed to eclipse, then $\sin{i} \approx 1$
and the age of the system is uniquely determined. Table~\ref{tab_modelfit}
lists the ages corresponding to this scenario, ranging from 2.5$^{+1.0}_{-0.7}$~Gyr to 3.2$^{+1.3}_{-0.9}$~Gyr
for the four models examined (uncertainties include the full range of possible
primary and secondary masses for which Eqn.~1 and the luminosity
constraints are satisfied).  The relatively small age uncertainties estimated in this scenario (25--60\%) are dominated by uncertainties in the
component luminosities, which could be measured from primary and secondary eclipse depths over a broad range of optical and infrared wavelengths (e.g., \citealt{sta06}).  
Such measurements are currently more feasible than resolved imaging measurements, as the tight separation inferred from the radial velocity orbit, $a\sin{i}$ = $a_1\frac{1+q}{q}\sin{i} \approx$ 0.4~AU (Bl08, assuming $q \approx 0.6$), implies a projected separation of $\lesssim$17~mas, below the diffraction limit of the Keck 10m telescope at near-infrared wavelengths.
Yet the scientifically valuable measurements possible in an eclipsing scenario must be tempered by this scenario's low probability.  For a maximum age of 10~Gyr, we can only constrain the inclination of the {\namesh} system to $i \gtrsim 53\degr$, and hence an eclipse probability of $\sim$0.3\%.

Regardless of whether {\namesh} is an eclipsing pair, 
determination of its orbit inclination and/or component masses is a necessary step for testing brown dwarf evolutionary models at late ages, specifically through agreement of system parameters with model isochrones (c.f.., \citealt{zap04}).  The power of such a test is the long lever arm of time
provided by older field binaries such as  {\namesh}, resulting in large differences in luminosities and effective temperatures for the more common near equal-mass systems (e.g., \citealt{all07}).
There may in fact be many such systems to exploit in this manner.
Simulations by Bu08 of brown dwarf pairs in the vicinity of the Sun predict that 12-25\% of 
all M8--L5 dwarf binaries have 
component spectral types that can be inferred from unresolved, near-infrared 
spectroscopy using the 
method outlined in \citet{me0805}.  Perhaps as many as 50\% of these systems may be
short-period radial velocity variables \citep{max05,bas06}.  Identification and follow-up of these systems
would complement the evolutionary model tests currently provided by younger systems
\citep{zap04,sta06,liu08,dup09} and would more robustly address
uncertainties associated with low-temperature light-element fusion, interior thermal transport,
and substellar interior structure as their effects are compounded over time.

\acknowledgements

The authors thank I.\ Baraffe, A.\ Burrows, M.\ Marley and D.\ Saumon for making electronic versions of their evolutionary models available; and M.\ Liu for identifying the
roundoff errors in the $M_{bol}$/spectral type relation in \citet{meltbinary}.
We also thank our referee, K.\ Luhman, for his helpful critique of the original manuscript.
CB acknowledges support from the Harvard Origins of Life Initiative. 
This publication has made use of the VLM Binaries Archive maintained by Nick Siegler at \url{http://www.vlmbinaries.org}.

\clearpage

\clearpage

\begin{deluxetable}{lcccc}
\tabletypesize{\footnotesize}
\tablecaption{Mass and Age Constraints for the {\name} System. \label{tab_modelfit}}
\tablewidth{0pt}
\tablehead{
\colhead{Parameter} &
\colhead{TUCSON} &
\colhead{COND} &
\colhead{SM08} &
\colhead{SM08} \\
\colhead{} &
\colhead{cloudless} &
\colhead{cloudless} &
\colhead{cloudless} &
\colhead{cloudy} \\
}
\startdata
Minimum age (Gyr) & 1.7 & 2.2 & 2.2 & 2.0 \\
Minimum $\sin{i}$\tablenotemark{a} & 0.80 (53$\degr$) & 0.82 (55$\degr$) & 0.83 (56$\degr$) & 0.86 (59$\degr$) \\
M$_1$ ({\msun})\tablenotemark{a} & 0.082--0.088 & 0.080--0.088 & 0.082--0.088 & 0.082--0.088 \\
M$_2$ ({\msun})\tablenotemark{a} & 0.054--0.075 & 0.053--0.070 & 0.054--0.069 & 0.054--0.066 \\
M$_2$/M$_1$ & 0.62--0.89 & 0.60--0.87 & 0.61--0.84 & 0.61--0.80 \\
Age for $\sin{i} = 1$ (Gyr) & 2.5$^{+1.0}_{-0.7}$ & 3.2$^{+1.3}_{-0.9}$ & 3.0$^{+1.2}_{-0.8}$ & 2.8$^{+1.7}_{-0.9}$ \\
\enddata
\tablenotetext{a}{Minimum $\sin{i}$ and maximum masses and mass ratios assume that the age of {\namesh} is less than 10~Gyr.}
\end{deluxetable}

\end{document}